\documentclass[fleqn,12pt]{article}
\usepackage{amsmath,amssymb,epsfig,times,graphics}
\usepackage{amssymb}
\usepackage{amsmath}
\usepackage{CJK}
\usepackage{graphics}
\usepackage{graphicx}
\usepackage{dcolumn}
\usepackage{bm}
\usepackage{subfigure}
\usepackage{lineno}
\usepackage{multirow}
\usepackage{amsthm}
\begin{document}
\begin{flushleft}
{\Large
\textbf{Stochastic stem cell models with mutation: A comparison of asymmetric and symmetric divisions} 
}
\newline
\\
Zhijie Wu\textsuperscript{1},
Yuman Wang\textsuperscript{1},
Kun Wang\textsuperscript{1},
Da Zhou$^*$\textsuperscript{1},
\\
\bigskip
\textbf{1} School of Mathematical Sciences, Xiamen University,
Xiamen 361005, People's Republic of China
\bigskip

*zhouda@xmu.edu.cn (DZ)
\end{flushleft}

\begin{abstract}
In order to fulfill cell proliferation and differentiation through cellular hierarchy, stem cells can undergo either asymmetric or symmetric divisions. Recent studies
pay special attention to the effect of different modes of stem cell division on the lifetime risk of cancer, and report that symmetric division is more beneficial to
delay the onset of cancer. The fate uncertainty of symmetric division is considered to be the reason for the cancer-delaying effect. In this paper we compare asymmetric
and symmetric divisions of stem cells via studying stochastic stem cell models with mutations. Specially, by using rigorous mathematical analysis we find that both asymmetric and symmetric models show the same statistical average, but symmetric model shows higher fluctuation than asymmetric model. We further show that
the difference between the two models would be more remarkable for lower mutation rates. Our work quantifies the uncertainty of cell division and highlights the significance of stochasticity for distinguishing between different modes of stem cell division.
\end{abstract}

\section{Introduction}
%
%
%
In multicellular organisms, many fast renewing tissues are organized in a hierarchical architecture, which proposes
a directional cascade from tissue specific stem cells to more differentiated cell states
\cite{yatabe2001investigating,tumbar2004defining,dingli2007compartmental,mackey2001cell}.
Stem cells possess two major properties: self-renewal and differential potential. That is, not only are
stem cells capable of maintaining the size of stem cell population by cell proliferation, but they can also
generate more specialized cell types by cell differentiation \cite{mcculloch2005perspectives,obernier2018adult}.

Two mechanisms of stem cells have been found to coordinate its dual role of self-renewal and differentiation \cite{morrison2006asymmetric}.
The first is asymmetric cell division \cite{knoblich2008mechanisms,zhong2008neurogenesis}, i.e. stem cells divide asymmetrically to give
rise to two daughter cells with different fates: one is identical to the mother stem cell, the other differentiates
into a non-stem cell state. The second is symmetric cell division \cite{shen2004endothelial}, namely, stem cells either perform self-renewal
by producing two stem cells identical to their mother, or perform differentiation by generating two differentiated daughter
cells. In particular, when the chances for self-renewal and differentiation are fifty-fifty, statistically speaking,
symmetric cell division is equivalent to asymmetric cell division in the sense that the gain and loss of stem cells are
balanced. In this case, tissue homeostasis can still be achieved in population level.
Even though both asymmetric and symmetric division modes are capable of maintaining tissue homeostasis, it appears that these two
mechanisms are not randomly distributed in the biological world. For example, strong evidence for asymmetric division
has been found in invertebrate systems \cite{Watt2000Out}, and symmetric division is more common in mammals
than in invertebrates \cite{simons2011strategies}. Therefore, exploring the evolutionary implications of different modes of cell division
is of great interest and importance \cite{tomasetti2010role,hu2013age}.
In recent years, the effect of stem cell division patterns on the risk of cancer has received special attention in
theoretical biology \cite{dingli2007stochastic,dingli2007symmetric,shahriyari2013symmetric,mchale2014protective,yang2015role,shahriyari2015role,
wodarz2014dynamics,werner2011dynamics}. Cancer is generally acknowledged as an evolutionary process involving
the accumulation of genetic or epigenetic mutations \cite{merlo2006cancer}. Note that mutations mostly result from errors during
the process of DNA replication, division mode of stem cells is supposed to have a significant impact on the process of tumorigenesis.
Despite the complexity of cancer, theoreticians still provide some interesting and
insightful researches on this issue. In particular,
Dingli et al showed that mutations with higher probability of asymmetric division
could result in rapid expansion of mutant stem cells \cite{dingli2007symmetric}.
Shahriyari and Komarova compared asymmetric, symmetric, and mixed stem cell divisions, and showed that symmetric stem cell divisions
could help to delay the onset of cancer \cite{shahriyari2013symmetric}.
McHale and Lander also reported similar result that symmetric divisions of stem cell make mutation accumulation slowly
\cite{mchale2014protective}. These results indicates that symmetric division mode has a
more significant cancer-delaying effect than asymmetric divisions, and the fate uncertainty of stochastic symmetric division
of mutant stem cells could be the reason for the cancer-delaying effect \cite{shahriyari2013symmetric,wodarz2014dynamics}.

To further address this issue, we here present a comparative study of asymmetric and symmetric divisions via
establishing stochastic stem cell models with mutation.
By using rigorous mathematical analysis, we obtain the explicit expressions of expectation and variance for
both wild-type and mutant cells.
We find that even though both asymmetric and symmetric division models
have the same expectation, their variances are quite different. Namely, both models
show the same statistical average, but symmetric model shows higher fluctuation than asymmetric model. This is in line
with previous observation that even though both symmetric and asymmetric division are able
to maintain tissue homeostasis, symmetric division result in greater uncertainty than
asymmetric division. The difference of variance between the two models is shown to be more remarkable for lower mutation rates.
\section{Models}

In order to model cellular hierarchies driven by different stem cell division patterns, we employ a compartment model framework compose of stem cell (type $A$) and non-stem cell (type $B$) \cite{rodriguez2011evolutionary,michor2003stochastic,zhou2018bayesian}.
Initially there are $N$ wild-type stem cells in the population. Non-stem cells are produced by stem cell differentiations. When a stem cell divides, mutation happens with
probability $P_0$, whereby either one of the daughter cells becomes mutant at random. In this model there are four different cell types: wild-type stem cell $A_0$, mutant stem cell $A_1$, wild-type non-stem cell $B_0$ and mutant non-stem cell $B_1$. We will incorporate asymmetric and symmetric
division patterns into the model framework respectively.
\subsection{Model for asymmetric division}

For asymmetric division mode, the schematic representation of the model is present as follows
\begin{align}
\begin{split}
\left \{
\begin{array}{ll}
A_0\xrightarrow{\lambda\left(1-P_0\right)}A_0+B_0\\
A_0\xrightarrow{\frac{\lambda P_0}2}A_1+B_0\\
A_0\xrightarrow{\frac{\lambda P_0}2}A_0+B_1\\
A_1\xrightarrow{\lambda} A_1+B_1
\end{array}
\right.
\end{split}
\label{Model1}
\end{align}
whereby each wild-type cell $A_0$ performs asymmetric cell division with rate $\lambda$, i.e. the waiting time for each asymmetric cell division
event follows exponential distribution with parameter $\lambda$. When it happens,
$A_0$ can either perform asymmetric division without any mutation (first arrow), or perform asymmetric division with
one mutant daughter cell (second and third arrows). Mutant stem cell $A_1$ can also perform asymmetric division giving rise to $A_1$ and
$B_1$ (fourth arrow). Here we assume that the division rate of $A_1$ is the same as $A_0$, i.e. neutral selection \cite{williams2016identification}.
Let $S_{A_0}\left(t\right)$ and $S_{A_1}\left(t\right)$
be the cell numbers of $A_0$ and $A_1$ at time $t$ respectively. Note that asymmetric division keeps the population size
of stem cell compartment constant, i.e.  $\forall t\geq0, S_{A_{0}}(t)+S_{A_{1}}(t)=N$. The stochastic dynamics
of asymmetric division model is captured by the probability distribution of $S_{A_{0}}(t)$ whose Kolmogorov forward equation \cite{gardiner1994handbook} is given by
\begin{align}
\frac{d}{d t}\left\{P\left[S_{A_{0}}(t)=x\right]\right\}=&-x \cdot P\left[S_{A_{0}}(t)=x\right] \cdot \frac{\lambda P_{0}}{2}
\notag\\
&+(x+1) P\left[S_{A_{0}}(t)=x+1\right] \cdot \frac{\lambda P_{0}}{2}~~~~~(x=0,1,2,3 \cdots N)
\label{CME1}
\end{align}
Let
\begin{equation}
E\left[S_{A_{0}}(t)\right]=\sum_{x=0}^{N} x\cdot P\left[S_{A_{0}}(t)=x\right]
\end{equation}
be the expectation characterizing the statistical average of $S_{A_{0}}(t)$, and
\begin{equation}
Var\left[S_{A_0}\left(t\right)\right]=E\left[S_{A_0}^2\left(t\right)\right]-E\left[S_{A_0}\left(t\right)\right]^2
\end{equation}
be the variance of $S_{A_{0}}(t)$ characterizing the stochastic fluctuation of $S_{A_{0}}(t)$ around the average.
Similarly we can define the expectation and variance for $S_{A_{1}}(t)$.
We are interested in how the expectation and variance are changed to different models of stem cell division.
\subsection{Model for symmetric division}

For symmetric division, the schematic representation of the model becomes
\begin{align}
\begin{split}
\left \{
\begin{array}{ll}
A_0\xrightarrow{\frac\lambda2\left(1-P_0\right)}A_0+A_0\\
A_0\xrightarrow{\frac\lambda2\left(1-P_0\right)}B_0+B_0\\
A_0\xrightarrow{\frac\lambda2P_0}A_0+A_1\\
A_0\xrightarrow{\frac\lambda2P_0}B_0+B_1\\
A_1\xrightarrow{\frac\lambda2}A_1+A_1\\
A_1\xrightarrow{\frac\lambda2}B_1+B_1
\end{array}
\right.
\end{split}
\label{Model2}
\end{align}
whereby each wild-type stem cell $A_0$ can either do cell proliferation (first arrow) or cell
differentiation (second arrow). When mutation happens, one of the daughter cells becomes
mutant (third and fourth arrows). By neutrality assumption, mutant stem cell $A_1$
performs symmetric divisions with rate $\lambda$ (fifth and sixth arrows).
In contrast to asymmetric model whereby $S_{A_{0}}(t)+S_{A_{1}}(t)=N$ all the time, in symmetric
model the sum of $S_{A_{0}}(t)$ and $S_{A_{1}}(t)$ is not constant anymore. Their joint probability
distribution is captured by
\begin{align}
&\frac{d\left\{P\left[S_{A_{1}}(t)=x_{1}, S_{A_{0}}(t)=x_{0}\right]\right\}}{d t}
\notag\\
&=P\left[S_{A_{1}}(t)=x_{1}, S_{A_{0}}(t)=x_{0}-1\right]\cdot \frac{\lambda\left(1-P_{0}\right)\left(x_{0}-1\right)}{2}
 \notag\\
&+P\left[S_{A_{1}}(t)=x_{1}, S_{A_{0}}(t)=x_{0}+1\right]\cdot \frac{\lambda\left(x_{0}+1\right)}{2}
 \notag\\
&-P\left[S_{A_{1}}(t)=x_{1}, S_{A_{0}}(t)=x_{0}\right]\cdot \frac{\lambda x_{0}}{2}
\notag \\
&-P\left[S_{A_{1}}(t)=x_{1}, S_{A_{0}}(t)=x_{0}\right]\cdot \frac{\lambda\left(1-P_{0}\right) x_{0}}{2}
\notag\\
&+P\left[S_{A_{1}}(t)=x_{1}-1, S_{A_{0}}(t)=x_{0}\right] \cdot \frac{\lambda P_{0} x_{0}}{2}
\notag\\
&-P\left[S_{A_{1}}(t)=x_{1}, S_{A_{0}}(t)=x_{0}\right] \cdot \frac{\lambda x_{1}}{2}
\notag \\
&+P\left[S_{A_{1}}(t)=x_{1}-1, S_{A_{0}}(t)=x_{0}\right] \cdot \frac{\lambda\left(x_{1}-1\right)}{2}
\notag\\
&+P\left[S_{A_{1}}(t)=x_{1}+1, S_{A_{0}}(t)=x_{0}\right] \cdot \frac{\lambda\left(x_{1}+1\right)}{2}
\notag\\
&-P\left[S_{A_{1}}(t)=x_{1}, S_{A_{0}}(t)=x_{0}\right] \cdot\left[\frac{\lambda x_{1}}{2}+\frac{\lambda P_{0} x_{0}}{2}\right].
\label{CME2}
\end{align}
We can also define the expectation and variance for $S_{A_{0}}(t)$ and $S_{A_{1}}(t)$ respectively.
In what follows we will compare asymmetric and symmetric models via calculating their expectations and variances.

\section{Results}
\subsection{Comparison of expectation}

We first check the expectation of the two models.
The main result is present in Theorem \ref{Thm1} (see \ref{app1} for proof)
\newtheorem{theorem}{Theorem}
\begin{theorem}
Both asymmetric model Eq. \eqref{CME1} and symmetric model Eq. \eqref{CME2} have the same expectations as follows:
\begin{equation}
E\left[S_{A_{0}}(t)\right]=\sum_{x=0}^{N} x\cdot P\left[S_{A_{0}}(t)=x\right]=N \cdot e^{-\frac{\lambda P_{0} t}{2}}
\end{equation}
\begin{equation}
E\left[S_{A_{1}}(t)\right]=\sum_{x=0}^{N} x\cdot P\left[S_{A_{1}}(t)=x\right]=N-N \cdot e^{-\frac{\lambda P_{0} t}{2}}
\end{equation}
\label{Thm1}
\end{theorem}
The result indicates that, even though
the two models have different cell-generating mechanisms, in the population level they are equivalent to each other
in the sense that on statistical average their loss of wild-type stem cell or gain of mutant stem cells follow the same function relation.
In order to explain this result, let us check the cellular processes of the two models in more details.
In asymmetric model, the process $A_0\xrightarrow{\frac{\lambda P_0}2}A_1+B_0$ produces one mutant stem cell and losses
one wild-type stem cell in a single step (with rate $\frac{\lambda P_{0}}{2}$). In symmetric model, the loss and gain of
stem cell is more complicated in the agent-based level. However, on average $A_0\xrightarrow{\frac\lambda2\left(1-P_0\right)}A_0+A_0$
and $A_0\xrightarrow{\frac\lambda2\left(1-P_0\right)}B_0+B_0$ are balanced, maintaining the number of wild-type stem cell substantially constant.
Besides, $A_0\xrightarrow{\frac\lambda2P_0}A_0+A_1$ produces one mutant stem cell at rate $\frac{\lambda P_{0}}{2}$ and
$A_0\xrightarrow{\frac\lambda2P_0}B_0+B_1$ losses one wild-type stem cell at the same rate. In this way, symmetric
model realizes the loss and gain of stem cells in several steps instead of one single step, but statistically speaking is
equivalent to one single step asymmetric division.

Another feature revealed by Theorem \ref{Thm1} is that
on average wild-type stem cells will die out ($e^{-\frac{\lambda P_{0} t}{2}}\rightarrow 0$ as
$t\rightarrow \infty$) and mutant stem cells will eventually take over the whole population of stem cells.
This is actually due to the fact that wild-type stem cell can become mutant but mutant stem cell cannot
revert to wild-type state. This prediction is in line with the coarse-grained model by
\cite{wodarz2014dynamics} (see Chapter 9). Things might be changed if more complicated selection rules are taken into account
(e.g. evolutionary game \cite{nowak2004evolutionary,pacheco2014ecology}), but even for the simple case here,
the process to mutant fixation would be quite diverse due to stochasticity of different division modes.
In next section we will calculate the variance for different models and
we will see that asymmetric and symmetric model are quite different.
\subsection{Comparison of variance}

Let $Var_a\left[\cdot\right]$ and $Var_s\left[\cdot\right]$ be the variances of \textbf{\b{a}}symmetric and \textbf{\b{s}}ymmetric models respectively.
The main results are present in Theorems \ref{Thm2}
and \ref{Thm3} as follows:

\begin{theorem}
For asymmetric model Eq. \eqref{CME1}, the variance of wild-type stem cell number is given by
\begin{align}
Var_a\left[S_{A_{0}}(t)\right]=N \cdot e^{-\frac{\lambda P_{0}t}{2}}\cdot\left(1-e^{\frac{-\lambda P_{0} t}{2}}\right).
\end{align}
Note that $S_{A_{1}}(t)=N-S_{A_{0}}(t)$, the variance of mutant stem cell number is also given by
\begin{align}
Var_a\left[S_{A_{1}}(t)\right]=Var_a\left[S_{A_{0}}(t)\right]=N \cdot e^{-\frac{\lambda P_{0}t}{2}}\cdot\left(1-e^{\frac{-\lambda P_{0} t}{2}}\right).
\end{align}
\label{Thm2}
\end{theorem}
The proof is present in \ref{app2}.
\begin{theorem}
For symmetric model Eq. \eqref{CME2}, the variance of wild-type stem cell number is given by
\begin{equation}
\begin{aligned}
Var_s\left[S_{A_{0}}(t)\right]=N \cdot \frac{\left(2-P_{0}\right)}{P_{0}} \cdot e^{-\frac{\lambda P_{0} t}{2}}\cdot\left(1-e^{-\frac{\lambda P_{0} t}{2}}\right).
\end{aligned}
\end{equation}
The variance of mutant stem cell number is given by
\begin{equation}
\begin{aligned}
Var_s\left[S_{A_{1}}(t)\right]
=\frac{N\left(2-P_{0}\right)}{P_{0}}\cdot e^{-\frac{\lambda P_{0} t}{2}}\cdot\left(1-e^{-\frac{\lambda P_{0} t}{2}}\right)+N \lambda t \cdot\left[1-\left(2-P_{0}\right) \cdot e^{-\frac{\lambda P_{0} t}{2}}\right].
\end{aligned}
\end{equation}
\label{Thm3}
\end{theorem}
The proof is present in \ref{app3}.

\begin{figure}
\centering
\includegraphics[scale=0.7]{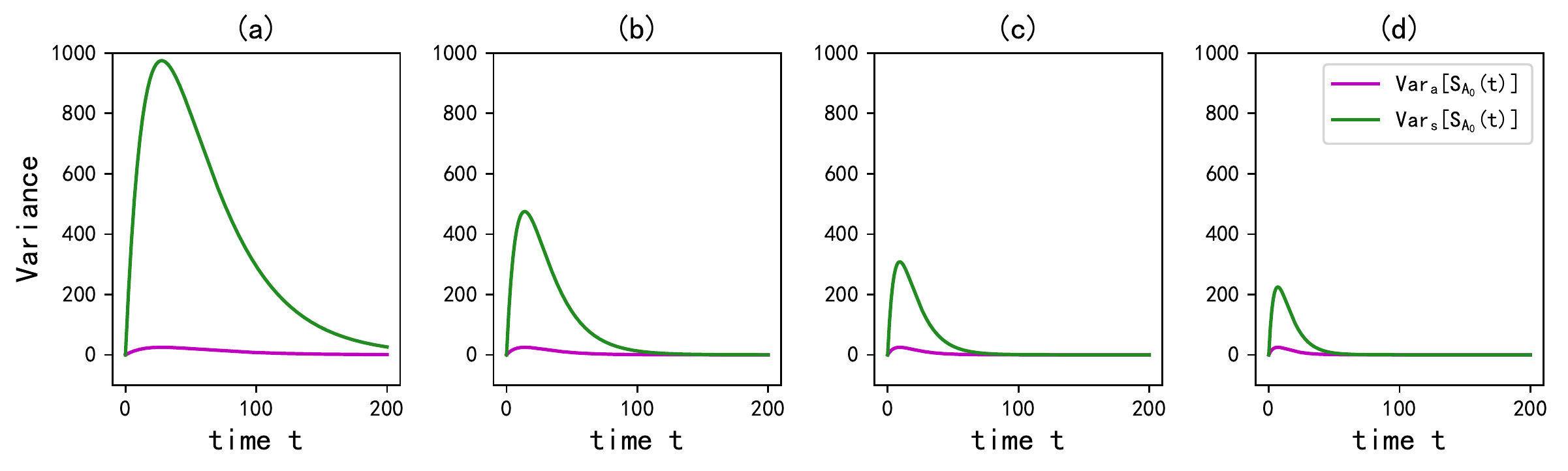}
\caption{The comparison between $Var_a\left[S_{A_{0}}(t)\right]$ and $Var_s\left[S_{A_{0}}(t)\right]$. From panel (a) to (d), the mutation probability $P_0=0.05, 0.1, 0.15, 0.2$.
The joint parameters are $N=100$, $\lambda=1$.}
\label{Fig1}
\end{figure}

For wild-type stem cells, it is easy to check that
\begin{equation}
\frac{Var_s\left[S_{A_{0}}(t)\right]}{Var_a\left[S_{A_{0}}(t)\right]}=\frac{\left(2-P_{0}\right)}{P_{0}}.
\end{equation}
Note that $P_0$ is mutation probability, i.e. $P_0\leq 1$, then we have
\begin{equation}
\frac{Var_s\left[S_{A_{0}}(t)\right]}{Var_a\left[S_{A_{0}}(t)\right]}\geq1.
\end{equation}
The condition for equality is $P_0=1$, and
the smaller $P_0$, the larger the difference between
$Var_s\left[S_{A_{0}}(t)\right]$ and $Var_a\left[S_{A_{0}}(t)\right]$
(Fig \ref{Fig1}). Note that the mutation probability
is generally very small ($P_0\ll1$), the distinction between $Var_a\left[S_{A_{0}}(t)\right]$ and $Var_s\left[S_{A_{0}}(t)\right]$ would be remarkable.
Recall that asymmetric and symmetric models have the same statistical average (Theorem \ref{Thm1}), variance rather than expectation could be more powerful to
differentiate between the two cell division mechanisms.

\begin{figure}
\centering
\includegraphics[scale=0.7]{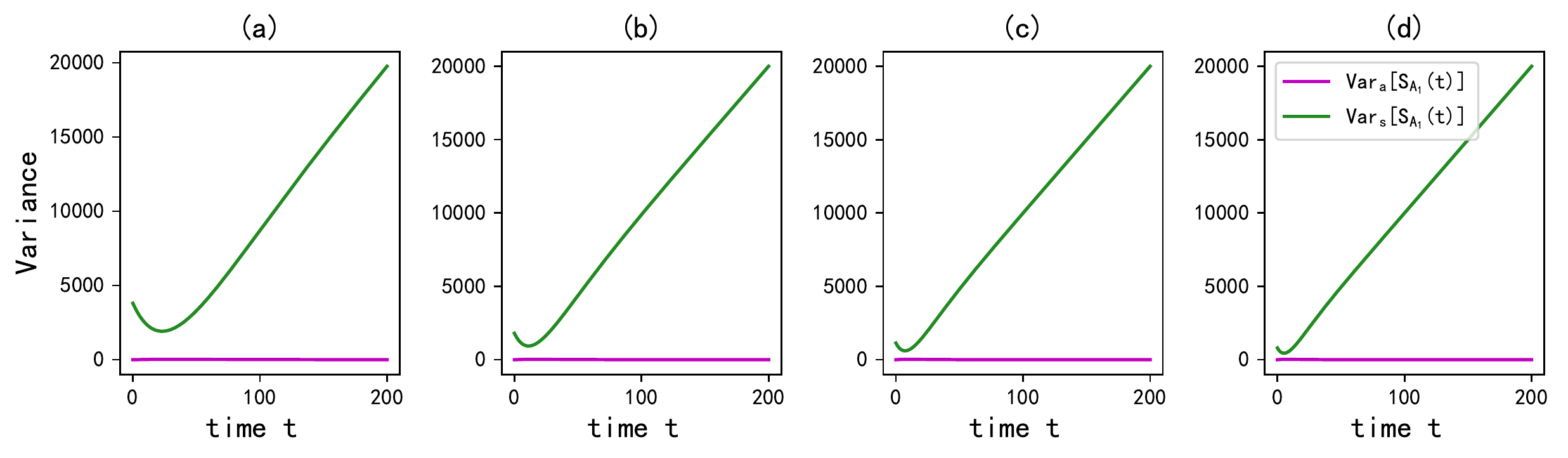}
\caption{The comparison between $Var_a\left[S_{A_{1}}(t)\right]$ and $Var_s\left[S_{A_{1}}(t)\right]$. From panel (a) to (d), the mutation probability $P_0=0.05, 0.1, 0.15, 0.2$.
The joint parameters are $N=100$, $\lambda=1$.}
\label{Fig2}
\end{figure}

Furthermore, an even more significant disparity between asymmetric and symmetric models is present in mutant stem cells.
From Fig \ref{Fig2}, we can see that $Var_a\left[S_{A_{1}}(t)\right]$ and $Var_s\left[S_{A_{1}}(t)\right]$ show
dramatically different trends. In contrast to $Var_a\left[S_{A_{1}}(t)\right]$ which tends to zero as time $t$ goes to infinity,
$Var_s\left[S_{A_{1}}(t)\right]\approx N\lambda t$ for large time $t$.
To explain it, recall that the expectation of wild-type stem cell will eventually die out (see Theorem \ref{Thm1}),
namely, for large time $t$, there are only mutant stem cells $A_1$ in the model. Note that
$S_{A_{1}}(t)$ either increases by one via $A_1\xrightarrow{\frac\lambda2}A_1+A_1$, or decrease by one via $A_1\xrightarrow{\frac\lambda2}B_1+B_1$,
so $S_{A_{1}}(t)$ can be regarded as a continuous-time symmetric random walk \cite{gardiner1994handbook,codling2008random}.
A standard property of the random walk model is that its variance
will increase linearly with time $t$. In this way, for mutant stem cells, very little fluctuation arises from asymmetric model as time goes by,
whereas the fluctuation arising from symmetric model is linearly mounting up with time (Fig. \ref{Fig2}). Still, this result quantitatively reveals
much higher uncertainty of symmetric division than asymmetric division.

\subsection{Comparison between asymmetric model and Moran-type symmetric model}

In previous section, we have compared the variances of asymmetric model and symmetric model, showing that
for both wild-type and mutant stem cells, symmetric division shows higher variance than asymmetric division. Note that the total number
of stem cells in asymmetric model remains unchanged, whereas the total stem cells number in symmetric model is variable,
so the extra uncertainties of symmetric model could come from the variability of the whole population size
instead of symmetric division pattern per se.
To check this issue, we present a Moran-type symmetric model whereby the total number of stem cells remains unchanged.
According to the birth and death events of stem cells, the six cellular processes in symmetric model Eq. \eqref{Model2} are classified into two classes:
\begin{itemize}
  \item Stem cell birth class: $A_0\xrightarrow{\frac\lambda2\left(1-P_0\right)}A_0+A_0$, $A_0\xrightarrow{\frac\lambda2P_0}A_0+A_1$, $A_1\xrightarrow{\frac\lambda2}A_1+A_1$.
  \item Stem cell death class: $A_0\xrightarrow{\frac\lambda2\left(1-P_0\right)}B_0+B_0$, $A_0\xrightarrow{\frac\lambda2P_0}B_0+B_1$, $A_1\xrightarrow{\frac\lambda2}B_1+B_1$.
\end{itemize}
Enlightened from the classical Moran process \cite{Moran1962},
we assume that at each update, a stem cell is chosen for death due to one of the three death events happening. For ensuring
the population size remains constant, one of the three birth events follows to happen.
In this way, a Moran-type symmetric model is captured by the following Kolmogorov forward equation
\begin{equation}
\begin{aligned}
&\frac{\left.d\left\{P [ S_{A_{0}}(t)=x\right]\right\}}{d t}\\
&=-\frac{\lambda\left(1-P_{0}\right)(N-x) x}{2 N}\cdot P\left[S_{A_{0}}(t)=x\right]\\
&-\frac{\lambda x\left[N-x\left(1-P_{0}\right)\right]}{2 N}\cdot P\left[S_{A_{0}}(t)=x\right]\\
&+\frac{\lambda\left(1-P_{0}\right)(N-x+1)(x-1)}{2 N}\cdot P\left[S_{A_{0}}(t)=x-1\right]\\
&+\frac{\lambda(x+1)\left[N-(x+1)\left(1-P_{0}\right)\right]}{2 N}\cdot P\left[S_{A_{0}}(t)=x+1\right],
\end{aligned}
\label{CME3}
\end{equation}
based on which we can calculate the expectation and variance. Let $Var_m\left[\cdot\right]$ be the variance of the \textbf{\b{M}}oran-type symmetric model.
The result is present in Theorem \ref{Thm4} (see \ref{app4} for the proof):
\begin{theorem}
For the Moran-type symmetric model Eq. \eqref{CME3}, the expectations of wild-type and mutant stem cells are given by
\begin{equation}
E\left[S_{A_{0}}(t)\right]=N \cdot e^{-\frac{\lambda P_{0} t}{2}},~~~E\left[S_{A_{1}}(t)\right]=N-N \cdot e^{-\frac{\lambda P_{0} t}{2}}.
\end{equation}
Note that $S_{A_{0}}(t)+S_{A_{1}}(t)=N$, their variances are the same and given by
\begin{equation}
\begin{aligned}
Var_m\left[S_{A_{0}}(t)\right]=Var_m\left[S_{A_{1}}(t)\right]&=\frac{\left(2-P_{0}\right) N^{2}}{N P_{0}+2\left(1-P_{0}\right)}\cdot e^{\frac{-\lambda P_{0}t}{2}}\\
&+\frac{N^{2} P_{0}(N-1)}{N P_{0}+2\left(1-P_{0}\right)}\cdot e^{-\left[\frac{N \lambda P_{0}+\lambda\left(1-P_{0}\right)}{N}\right] \cdot t}-N^{2}\cdot e^{-\lambda P_{0} t}.
\end{aligned}
\end{equation}
\label{Thm4}
\end{theorem}
From Theorem \ref{Thm4}, it is easy to see that the expectations of Moran-type symmetric model are the same as those of previous two models, namely,
they share the same statistical average.

\begin{figure}
\centering
\includegraphics[scale=0.7]{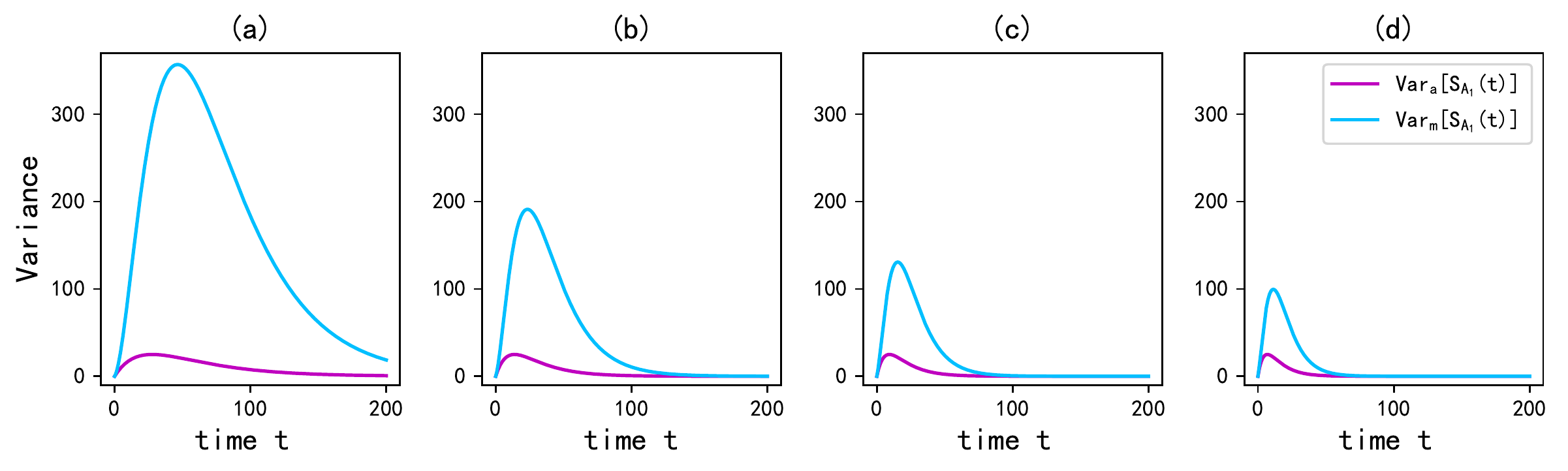}
\caption{The comparison between $Var_a\left[S_{A_{1}}(t)\right]$ and $Var_m\left[S_{A_{1}}(t)\right]$. From panel (a) to (d), the mutation probability $P_0=0.05, 0.1, 0.15, 0.2$.
The joint parameters are $N=100$, $\lambda=1$.}
\label{Fig3}
\end{figure}

For variance, compared to the symmetric model Eq. \eqref{CME2}, the Moran-type symmetric model removes the effect of the variability of the whole population size.
Namely, the fluctuation of the Moran-type symmetric model comes from the fate uncertainty of symmetric cell division per se.
Fig \ref{Fig3} shows that $Var_m\left[S_{A_{1}}(t)\right]$ is bounded and tends to zero instead of going to infinity as time goes by. Even so,
we can see that the Moran-type symmetric model
still shows larger variance than asymmetric model (Fig \ref{Fig3}). Mathematically we can prove that (see \ref{app5})
\begin{equation}
Var_m\left[S_{A_{1}}(t)\right]\geq Var_a\left[S_{A_{1}}(t)\right].
\label{inequ}
\end{equation}
The condition for equality is $P_0=1$.
The smaller $P_0$, the larger the difference between $Var_m\left[S_{A_{1}}(t)\right]$ and $Var_a\left[S_{A_{1}}(t)\right]$.
Hence Moran-type symmetric model generally show larger fluctuation around average than asymmetric model, especially for rare
mutation cases ($P_0\ll 1$).

\section{Conclusions}

In this study, we have explored how different cell division modes affect the statistical average (expectation) and the fluctuation around average
(variance) of the stochastic stem cell models with mutation. By using rigorous mathematical analysis, we have shown that asymmetric model,
symmetric model and Moran-type symmetric model have the same expectations. However, their variances are quite different. Symmetric divisions
(both symmetric model and Moran-type symmetric model) show larger variance than asymmetric division model. The fate uncertainty of symmetric division
(either cell proliferation or differentiation) and the variability of the total population size are the major sources of the variance
arising from symmetric division models.

Our results reveal the importance of stochasticity for distinguishing between different division patterns. Note that the deterministic dynamics of asymmetric
and symmetric divisions are the same, namely, it is quite impossible to identify cell division mode only based on average measurements.
More attention should be paid to stochastic model, which is not an alternative to deterministic model but a more complete description
\cite{qian2002concentration,jiang2017phenotypic,niu2015phenotypic}. As a supplement to statistical average, the fluctuation around average
has been proved to be very important information for model comparison and selection \cite{zhou2018bayesian}. Besides, since variance
is very sensitive to division pattern and mutation rate, it can also be used to develop efficient parameter estimation method
combining the stochastic stem cell model and high-resolution experimental data, which would be of great value in future researches.

\section*{Acknowledgements}

This work is supported by the National Natural Science Foundation of China (Grant
No. 11971405), the Fundamental Research Funds for the Central Universities in China (Grant No. 20720180005).

\appendix

\section{Proof of Theorem \ref{Thm1}}
\label{app1}

For asymmetric model Eq. \eqref{CME1}, note that $E\left[S_{A_{1}}(t)\right]=N-E\left[S_{A_{0}}(t)\right]$, it
is sufficient to prove $E\left[S_{A_{0}}(t)\right]=N \cdot e^{-\frac{\lambda P_{0} t}{2}}$. Considering the
derivative of $E\left[S_{A_{0}}(t)\right]$

\begin{align}
 &\frac{d\left[E\left(S_{A_{0}}(t)\right)\right]}{d t} =\frac{d\left(\sum_{x=0}^{N} x \cdot P\left[S_{A_{0}}(t)=x\right]\right)}{d t}=\sum_{x=0}^{N} x\cdot \frac{d\left\{P\left[S_{A_{0}}(t)=x\right]\right\}}{d t}
\end{align}
and plugging Eq. \eqref{CME1} we have
\begin{align}
&\frac{d\left[ E\left(S_{A_{0}}(t)\right)\right]}{d t}=\sum_{x=0}^{N} x \cdot \frac{d\left\{P\left[S_{A_{0}}(t)=x\right]\right\}}{d t}
\notag\\
&=\sum_{x=1}^{N-1} x \cdot \frac{d\left\{P\left[S_{A_{0}}(t)=x\right]\right\}}{d t}+0 \cdot \frac{d\left\{P\left[S_{A_{0}}(t)=0\right]\right\}}{d t}+N \cdot \frac{d\left\{P\left[S_{A_{0}}(t)=N\right]\right\}}{d t}
\notag\\
&=\sum_{x=1}^{N-1}\left\{-x^{2} \cdot P\left[S_{A_{0}}(t)=x\right]+(x+1-1)(x+1)\cdot P\left[S_{A_{0}}(t)=x+1\right]\right\} \cdot \frac{\lambda P_{0}}{2}
\notag\\
&-N^{2} \cdot\left\{P\left[S_{A_{0}}(t)=N\right]\right\} \cdot \frac{\lambda P_{0}}{2}
\notag\\
&=-\sum_{x=1}^{N-1} x^{2} \cdot \frac{\lambda P_{0}}{2}\cdot P\left[S_{A_{0}}(t)=x\right]+\sum_{x=1}^{N} x^{2} \cdot \frac{\lambda P_{0}}{2}\cdot P\left[S_{A_{0}}(t)=x\right]
\notag\\
&-\sum_{x=1}^{N} x \cdot \frac{\lambda P_{0}}{2}\cdot P\left[S_{A_{0}}(t)=x\right]-N^{2} \cdot\ P\left[S_{A_{0}}(t)=N\right] \cdot \frac{\lambda P_{0}}{2}
\notag\\
&=N^{2} \cdot \frac{\lambda P_{0}}{2} \cdot P\left[S_{A_{0}}(t)=N\right]-\sum_{x=1}^{N} x \cdot \frac{\lambda P_{0}}{2}\cdot P\left[S_{A_{0}}(t)=x\right]
\notag\\
&-N^{2} \cdot P\left[S_{A_{0}}(t)=N\right] \cdot \frac{\lambda P_{0}}{2}=-E\left[S_{A_{0}}(t)\right]\cdot \frac{\lambda P_{0}}{2}
\end{align}
Namely,
\begin{align}
\frac{d\left\{ E\left[S_{A_{0}}(t)\right]\right\}}{d t}=-E\left[S_{A_{0}}(t)\right]\cdot \frac{\lambda P_{0}}{2}.
\end{align}
Given the initial condition $E\left[S_{A_{0}}(0)\right]=N$, we obtain the solution that
\begin{align}
E\left[S_{A_{0}}(t)\right]=N \cdot e^{-\frac{\lambda P_{0} t}{2}}.
\end{align}

For the symmetric model, based on the joint distribution given by Eq. \eqref{CME2}, we have
\begin{align}
&\frac{d\left\{P\left[S_{A_{0}}(t)=x_{0}\right]\right\}}{d t}=\sum_{x_{1}=0}^{+\infty} \frac{d\left\{P\left[S_{A_{1}}(t)=x_{1}, S_{A_{0}}(t)=x_{0}\right]\right\}}{d t}
\notag\\
&=\sum_{x_{1}=1}^{+\infty} \frac{d\left\{P\left[S_{A_{1}}(t)=x_{1}, S_{A_{0}}(t)=x_{0}\right]\right\}}{d t}+ \frac{d\left\{P\left[S_{A_{1}}(t)=0, S_{A_{0}}(t)=x_{0}\right]\right\}}{d t}
\notag\\
&=P\left[S_{A_{0}}(t)=x_{0}-1\right] \cdot \frac{\lambda\left(1-P_{0}\right)\left(x_{0}-1\right)}{2}
\notag\\
&+P\left[S_{A_{0}}(t)=x_{0}+1\right] \cdot \frac{\lambda\left(x_{0}+1\right)}{2}-P\left[S_{A_{0}}(t)=x_{0}\right] \cdot \frac{\lambda\left(2-P_{0}\right) x_{0}}{2}
\notag\\
&+P\left[S_{A_{0}}(t)=x_{0}\right] \cdot \frac{\lambda P_{0} x_{0}}{2}
\end{align}
Then considering the derivative of $E \left[ S_{A_{0}}(t)\right]$
\begin{align}
&\frac{\left.d\left\{E [ S_{A_{0}}(t)\right]\right\}}{d t}
\notag\\
 &=\sum_{x=1}^{+\infty} x \cdot P\left[S_{A_{0}}(t)=x-1\right] \cdot \frac{\lambda\left(1-P_{0}\right)(x-1)}{2}
\notag\\
&+\sum_{x=1}^{+\infty} x \cdot P\left[S_{A_{0}}(t)=x+1\right] \cdot \frac{\lambda(x+1)}{2}-\sum_{x=1}^{+\infty} x \cdot\left[S_{A_{0}}(t)=x\right] \cdot \frac{\lambda\left(2-P_{0}\right) x}{2}
 \notag\\
&=\sum_{x=1}^{+\infty}(x-1+1) \cdot P\left[S_{A_{0}}(t)=x-1\right] \cdot \frac{\lambda\left(1-P_{0}\right)(x-1)}{2}
\notag \\
&+\sum_{x=1}^{+\infty}(x+1-1) \cdot P\left[S_{A_{0}}(t)=x+1\right] \cdot \frac{\lambda(x+1)}{2}
\notag\\
&-\sum_{x=1}^{+\infty} x \cdot\left[S_{A_{0}}(t)=x\right] \cdot \frac{\lambda\left(2-P_{0}\right) x}{2}
\notag\\
&=\sum_{x=1}^{+\infty} x^{2} \cdot P\left[S_{A_{0}}(t)=x\right] \cdot \frac{\lambda\left(1-P_{0}\right)}{2}
+E\left[S_{A_{0}}(t)\right] \cdot \frac{\lambda\left(1-P_{0}\right)}{2}
\notag\\
&=-E\left[S_{A_{0}}(t)\right] \cdot \frac{\lambda P_{0}}{2}
\end{align}
By solving the above equation we have
\begin{align}
E\left[S_{A_{0}}(t)\right]=N \cdot \mathrm{e}^{-\frac{\lambda P_{0} t}{2}}.
\end{align}\\
We now calculate $E\left[S_{A_{1}}(t)\right]$ for symmetric model. Note that in the symmetric model, $S_{A_{0}}(t)+S_{A_{1}}(t)$
is not constant. Based on Eq. \eqref{CME2} we have

\begin{equation}
\begin{aligned}
&\frac{d\left\{P\left[S_{A_{1}}(t)=x_{1}\right]\right\}}{d t}=\sum_{x_{0}=0}^{+\infty} \frac{d\left\{P \left[S_{A_{1}}(t)=x_{1}, S_{A_{0}}(t)=x_{0}\right]\right\}}{d t}\\
&=- \frac{\lambda P_{0}}{2}\cdot E\left[S_{A_{0}}(t)=x_{0} \mid S_{A_{1}}(t)=x_{1}\right]\cdot P\left[S_{A_{1}}(t)=x_{1}\right] \\
&+\frac{\lambda P_{0}}{2}\cdot E\left[S_{A_{0}}(t)=x_{0} \mid S_{A_{1}}(t)=x_{1}-1\right]\cdot P\left[S_{A_{1}}(t)=x_{1}-1\right]\\
&-P\left[S_{A_{1}}(t)=x_{1}\right] \cdot \lambda x_{1} \\
&+P\left[S_{A_{1}}(t)=x_{1}-1\right]\cdot \frac{\lambda\left(x_{1}-1\right)}{2}+P\left[S_{A_{1}}(t)=x_{1}+1\right] \cdot \frac{\lambda\left(x_{1}+1\right)}{2}.
\end{aligned}
\end{equation}
Then the derivative of $E\left[S_{A_{1}}(t)\right]$ is given by
\begin{equation}
\begin{aligned}
&\frac{d\left\{E\left[S_{A_{1}}(t)\right]\right\}}{d t}\\
&=+\frac{\lambda P_{0}}{2}\cdot \sum_{x_{1}=0}^{+\infty} E\left[S_{A_{0}}(t) \mid S_{A_{1}}(t)=x_{1}\right]\cdot P\left[S_{A_{1}}(t)=x_{1}\right] \\
 &-\sum_{x_{1}=1}^{+\infty} \lambda{x_{1}}\cdot P\left[S_{A_{1}}(t)=x_{1}\right]  \\
 &+\sum_{x_{1}=1}^{+\infty}\frac{\lambda}{2} x_{1}\cdot P\left[S_{A_{1}}(t)=x_{1}\right]+\sum_{x_{1}=1}^{+\infty}\frac{\lambda}{2}x_{1}\cdot P\left[S_{A_{1}}(t)=x_{1}\right] \\
&+\sum_{x_{1}=1}^{+\infty}\frac{\lambda }{2} x_{1}^{2}\cdot P\left[S_{A_{1}}(t)=x_{1}\right]-\sum_{x_{1}=1}^{+\infty}\frac{\lambda }{2}{x_{1}}^2 \cdot P\left[S_{A_{1}}(t)=x_{1}\right] \\
&=\frac{\lambda P_{0}}{2}\cdot E\left[S_{A_{0}}(t)\right].
\end{aligned}
\end{equation}
Solving the above equation as follows completes the proof
\begin{equation}
E\left[S_{A_{1}}(t)\right]=N \cdot\left(1-e^{-\frac{\lambda P_{0}t}{2}}\right)
\end{equation}

\section{Proof of Theorem \ref{Thm2}}
\label{app2}

In order to calculate $Var\left[S_{A_{0}}(t)\right]$, it is sufficient to calculate $E\left[S_{A_0}^{2}(t)\right]$. Based on Eq. \eqref{CME1} we have
\begin{align}
&\frac{d\left\{E\left[S_{A 0}^{2}(t)\right]\right\}}{d t}
\notag\\
&=\sum_{x=1}^{N}-\frac{\lambda P_{0}}{2}x^{3}\cdot P\left[S_{A_{0}}(t)=x\right] +\sum_{x=1}^{N} \frac{\lambda P_{0}}{2}x^{3} \cdot P\left[S_{A_{0}}(t)=x\right]
\notag\\
&-\sum_{x=1}^{N} \lambda P_{0} x^{2} \cdot P\left[S_{A_{0}}(t)=x\right]+\sum_{x=1}^{N}\frac{\lambda P_{0}}{2} x\cdot P\left[S_{A_{0}}(t)=x\right]
\notag\\
&=-\lambda P_{0}\cdot E\left[S_{A_{0}}^{2}(t)\right] +\frac{\lambda P_{0}}{2}\cdot E\left[S_{A_{0}}(t)\right]
 \end{align}
By using the method of variation of constant and the fact that
\begin{align}
Var\left[S_{A_{0}}(t)\right]=E\left[S_{A_{0}}^{2}(t)\right]-\left\{E\left[S_{A_{0}}(t)\right]\right\}^{2},
\end{align}
we have
\begin{align}
Var\left[S_{A_{0}}(t)\right] &=N \cdot e^{-\frac{\lambda P_{0}t}{2}}-N \cdot e^{-\lambda P_{0} t}
 \notag\\
&=N \cdot e^{-\frac{\lambda P_{0}t}{2}}\cdot\left(1-e^{\frac{-\lambda P_{0} t}{2}}\right)
\end{align}
Note that
\begin{align}
S_{A_{0}}(t)+S_{A_{1}}(t)=N,
\end{align}
then we have
\begin{align}
Var\left[S_{A_{1}}(t)\right]=Var\left[S_{A_{0}}(t)\right]=N \cdot e^{-\frac{\lambda P_{0} t}{2}} \cdot\left(1-e^{-\frac{\lambda P_{0} t}{2}}\right)
\end{align}

\section{Proof of Theorem \ref{Thm3}}
\label{app3}

We first calculate $Var\left[S_{A_{0}}(t)\right]$. Note that
\begin{align}
Var\left[S_{A_{0}}(t)\right]=E\left[S_{A_{0}}^{2}(t)\right]-\left\{E\left[S_{A_{0}}(t)\right]\right\}^{2},
\label{Vd}
\end{align}
it is sufficient to calculate $E\left[S_{A_{0}}^{2}(t)\right]$.
\begin{align}
\frac{d\left\{E\left[S_{A_{0}}^{2}(t)\right]\right\}}{d t}
\notag\\
=\text{ }\sum_{x=0}^{+\infty} & x^{2}\cdot\frac{P\left[ S_{A_{0}}(t)=x\right]}{d t}
\notag\\
=\text{ }\sum_{x=1}^{+\infty} & \frac{(x-1+1)^{2}(x-1) \lambda\left(1-P_{0}\right)}{2} \cdot P\left[S_{A_{0}}(t)=x-1\right]
\notag\\
 +\text{ }\sum_{x=1}^{+\infty} & \frac{(x+1-1)^{2} \lambda(x+1)}{2} \cdot P\left[S_{A_{0}}(t)=x+1\right]
\notag\\
-\text{ }\sum_{x=1}^{+\infty}&  \frac{x^{3} \lambda\left(2-P_{0}\right)}{2} \cdot P\left[S_{A_{0}}(t)=x\right]
\notag \\
 =\text{ }\sum_{x=1}^{+\infty} & \frac{x^{3} \lambda\left(1-P_{0}\right)}{2} \cdot P\left[S_{A_{0}}(t)=x\right]
\notag \\
 +\text{ }\sum_{x=1}^{+\infty} & \frac{2 x^{2} \lambda\left(1-P_{0}\right)}{2} \cdot P\left[S_{A_{0}}(t)=x\right]
\notag\\
+\text{}\sum_{x=1}^{+\infty}& \frac{x \lambda\left(1-P_{0}\right)}{2} \cdot P\left[S_{A_{0}}(t)=x\right]
\notag \\
 +\text{ }\sum_{x=2}^{+\infty}& \frac{\lambda x^{3}}{2} \cdot P\left[S_{A_{0}}(t)=x\right]
\notag\\
-\text{ }\sum_{x=2}^{+\infty}& \frac{2 \lambda x^{2}}{2} \cdot P\left[S_{A_{0}}(t)=x\right]+\sum_{x=2}^{+\infty}\frac{\lambda x}{2} \cdot P\left[S_{A_{0}}(t)=x\right]
\notag\\
 -\text{ }\sum_{x=1}^{+\infty} & \frac{x^{3} \lambda\left(1-P_{0}\right)}{2} \cdot P\left[S_{A_{0}}(t)=x\right]-\sum_{x=1}^{+\infty} \cdot \frac{x^{3} \lambda}{2} \cdot P\left[S_{A_{0}}(t)=x\right]
\notag\\
=\text{ }&-\lambda P_{0}\cdot E \left[S_{A_{0}}^{2}(t)\right]+E \left[S_{A_{0}}(t)\right)]\cdot\frac{\lambda\left(2-P_{0}\right)}{2}
\end{align}
The above equation is solved using the method of variation of constant, and by using Eq. \eqref{Vd} we have
\begin{equation}
\begin{aligned}
Var \left[S_{A_{0}}(t)\right)]&=N \cdot \frac{\left(2-P_{0}\right)}{P_{0}} \cdot e^{-\frac{\lambda P_{0} t}{2}}-N \cdot \frac{\left(2-P_{0}\right)}{P_{0}} \cdot e^{-\lambda P_{0} t} \\
&=N \cdot \frac{\left(2-P_{0}\right)}{P_{0}} \cdot e^{-\frac{\lambda P_{0} t}{2}}\cdot\left(1-e^{-\frac{\lambda P_{0} t}{2}}\right) \\
&=N \cdot \frac{\left(2-P_{0}\right)}{P_{0}} \cdot e^{-\frac{\lambda P_{0} t}{2}}-N \cdot \frac{\left(2-P_{0}\right)}{P_{0}} \cdot e^{-\lambda P_{0} t}.
\end{aligned}
\end{equation}
We next calculate $Var\left[S_{A_{1}}(t)\right]$.
\begin{equation}
\begin{aligned}
&\frac{d\left\{Var\left[S_{A_{1}}(t)\right]\right\}}{dt}= \frac{d\left\{E\left[S_{A_{1}}^{2}(t)\right]-E^{2}\left[S_{A_{1}}(t)\right]\right\}}{d t}\\
&=\frac{d\left\{E\left[S_{A_{1}}^{2}(t)\right]\right\}}{d t}-2 E\left[S_{A_{1}}(t)\right] \cdot \frac{d\left\{E\left[S_{A_{1}}(t)\right]\right\}}{d t}\\
&=\lambda P_{0}\cdot \sum_{x_{1}=1}^{+\infty} \sum_{x_{0}=1}^{+\infty} x_{1} x_{0}\cdot P\left[S_{A_{0}}(t)=x_{0}, S_{A_{1}}(t)=x_{1}\right]\\
&+\frac{\lambda P_{0}}{2}\cdot E\left[S_{A_{0}}(t)\right]+\lambda \cdot E\left[S_{A_{1}}(t)\right] \\
& =\lambda P_{0}\cdot E\left[S_{A_{0}}(t) \cdot S_{A_{1}}(t)\right]+\frac{\lambda P_{0}}{2}\cdot E\left[S_{A_{0}}(t)\right]+N \lambda
\\&-\lambda\cdot E\left[S_{A_{0}}(t)\right]-N \lambda P_{0}\cdot E\left[S_{A_{0}}(t)\right]+\lambda P_{0}\cdot E^{2}\left[S_{A_{0}}(t)\right] \\
&= \lambda P_{0}\cdot E\left[S_{A_{0}}(t) \cdot S_{A_{1}}(t)\right]+\left(\frac{\lambda P_{0}}{2}-\lambda-N \lambda P_{0}\right) \cdot E\left[S_{A_{0}}(t)\right]\\
&+\lambda P_{0}\cdot E^{2}\left[S_{A_{0}}(t)\right]+N \lambda.
\end{aligned}
\end{equation}
Hence it is sufficient to calculate E $\left[S_{A_{0}}(t) \cdot S_{A_{1}}(t)\right]$.
\begin{equation}
\begin{aligned}
 &\frac{d\left\{E\left[S_{A_{0}}(t) \cdot S_{A_{1}}(t)\right]\right\}}{d t}\\
 &=\frac{d\left\{\sum_{x_{1}=0}^{+\infty} \sum_{x_{0}=0}^{+\infty} x_{1} x_{0} P\left[S_{A_{0}}(t)=x_{0}, S_{A_{1}}(t)=x_{1}\right]\right\}}{d t} \\ &=\sum_{x_{1}=1}^{\infty} \sum_{x_{0}=1}^{\infty} x_{1} x_{0} \cdot \frac{d\left\{P\left[S_{A_{0}}(t)=x_{0}, S_{A_{1}}(t)=x_{1}\right]\right\}}{d t}
 \end{aligned}
\end{equation}
Note that
\begin{equation}
\begin{aligned}
&\sum_{x_{1}=1}^{+\infty} \sum_{x_{0}=1}^{+\infty} x_{1} x_{0}\cdot \frac{\left.d\left\{P [ S_{A_{1}}(t)=x_{1}, S_{A_{0}}(t)=x_{0}\right]\right\}}{d t}\\
&=\sum_{x_{1}=1}^{+\infty} \sum_{x_{0}=1}^{+\infty} x_{1} x_{0}\cdot P\left[S_{A_{1}}(t)=x_{1}, S_{A_{0}}(t)=x_{0}-1\right] \cdot \frac{\lambda\left(1-P_{0}\right)\left(x_{0}-1\right)}{2}\\
&+\sum_{x_{1}=1}^{+\infty} \sum_{x_{0}=1}^{+\infty} x_{1} x_{0}\cdot P\left[S_{A_{1}}(t)=x_{1}, S_{A_{0}}(t)=x_{0}+1\right] \cdot \frac{\lambda\left(x_{0}+1\right)}{2}\\
&-\sum_{x_{1}=1}^{+\infty} \sum_{x_{0}=1}^{+\infty} x_{1} x_{0}\cdot P\left[S_{A_{1}}(t)=x_{1}, S_{A_{0}}(t)=x_{0}\right] \cdot \frac{\lambda x_{0}}{2}\\
&-\sum_{x_{1}=1}^{+\infty} \sum_{x_{0}=1}^{+\infty} x_{1} x_{0}\cdot P\left[S_{A_{1}}(t)=x_{1}, S_{A_{0}}(t)=x_{0}\right] \cdot \frac{\lambda\left(1-P_{0}\right) x_{0}}{2}\\
&+\sum_{x_{1}=1}^{+\infty} \sum_{x_{0}=1}^{+\infty} x_{1} x_{0} \cdot P\left[S_{A_{1}}(t)=x_{1}-1, S_{A_{0}}(t)=x_{0}\right] \cdot \frac{\lambda P_{0} x_{0}}{2}\\
&-\sum_{x_{1}=1}^{+\infty} \sum_{x_{0}=1}^{+\infty} x_{1} x_{0}\cdot P\left[S_{A_{1}}(t)=x_{1}, S_{A_{0}}(t)=x_{0}\right] \cdot \frac{\lambda x_{1}}{2}\\
&+\sum_{x_{1}=1}^{+\infty} \sum_{x_{0}=1}^{+\infty} x_{1} x_{0}\cdot P\left[S_{A_{1}}(t)=x_{1}-1, S_{A_{0}}(t)=x_{0}\right] \cdot \frac{\lambda\left(x_{1}-1\right)}{2}\\
&+\sum_{x_{1}=1}^{+\infty} \sum_{x_{0}=1}^{+\infty} x_{1} x_{0}\cdot P\left[S_{A_{1}}(t)=x_{1}+1, S_{A_{0}}(t)=x_{0}\right] \cdot \frac{\lambda\left(x_{1}+1\right)}{2}\\
&-\sum_{x_{1}=1}^{+\infty} \sum_{x_{0}=1}^{+\infty} x_{1} x_{0}\cdot P\left[S_{A_{1}}(t)=x_{1}, S_{A_{0}}(t)=x_{0}\right] \cdot\left[\frac{\lambda x_{1}}{2}+\frac{\lambda P_{0} x_{0}}{2}\right]
\end{aligned}
\end{equation}
we have
\begin{equation}
\begin{aligned}
&\frac{\left.d\left\{E [ S_{A_{1}}(t) \cdot S_{A_{0}}(t)\right]\right\}}{d t}\\
&= \sum_{x_{1}=0}^{+\infty} \sum_{x_{0}=0}^{+\infty} x_{1} x_{0} \cdot\frac{\left.d\left\{P [S_{A_{1}}(t)=x_{1}, S_{A_{0}}(t)=x_{0}\right]\right\}}{d t} \\
&=\frac{\lambda P_{0}}{2}\cdot E\left[S_{A_{0}}^{2}(t)\right]-\frac{\lambda P_{0}}{2}\cdot \sum_{x_{1}=0}^{+\infty} \sum_{x_{0}=0}^{+\infty} x_{1} x_{0}\cdot P\left[S_{A_{1}}(t)=x_{1}, S_{A_{0}}(t)=x_{0}\right] \\
&=\frac{\lambda P_{0}}{2}\cdot E\left[S_{A_{0}}^{2}(t)\right]-\frac{\lambda P_{0}}{2}\cdot E\left[S_{A_{1}}(t) \cdot S_{A_{0}}(t)\right]
\end{aligned}
\end{equation}
Given
\begin{equation}
\begin{aligned}
Var \left[S_{A_{0}}(t)\right]=N \cdot \frac{\left(2-P_{0}\right)}{P_{0}} \cdot e^{-\frac{\lambda P_{0} t}{2}}\cdot\left(1-e^{-\frac{\lambda P_{0} t}{2}}\right)
\end{aligned}
\end{equation}
and
\begin{equation}
\begin{aligned}
E\left[S_{A_{0}}(t)\right]=N \cdot e^{-\frac{\lambda P_{0} t}{2}},
\end{aligned}
\end{equation}
we have
\begin{equation}
\begin{aligned}
E\left[S_{A_{0}}^{2}(t)\right] &=E^{2}\left[S_{A_{0}}(t)\right]+Var\left[S_{A_{0}}(t)\right]
 \\
&=\left(N^{2}-N \cdot \frac{2-P_{0}}{P_{0}}\right) \cdot e^{-\lambda P_{0} t}+N \cdot \frac{2-P_{0}}{P_{0}} \cdot e^{-\frac{\lambda P_{0} t}{2}}
\end{aligned}
\end{equation}
It turns out that
\begin{equation}
\begin{aligned}
\frac{\left.d\left\{E[S_{A_{1}}(t) \cdot S_{A_{0}}(t)\right]\right\}}{d t}& \\
=\frac{\lambda P_{0}}{2}& \cdot\left(N^{2}-N\cdot\frac{2-P_{0}}{P_{0}}\right) \cdot e^{-\lambda P_{0} t}\\
+N \cdot &\frac{\lambda\left(2-P_{0}\right)}{2} \cdot e^{-\frac{\lambda P_{0} t}{2}}-\frac{\lambda P_{0}}{2}\cdot E\left[S_{A_{1}}(t) \cdot S_{A_{0}}(t)\right].
\end{aligned}
\end{equation}
Its solution is given by
\begin{equation}
\begin{aligned}
E\left[S_{A_{0}}(t) \cdot S_{A_{1}}(t)\right]&
=\left(N^{2}-N \cdot \frac{2-P_{0}}{P_{0}}\right) \cdot e^{-\frac{\lambda P_{0} t}{2}} \cdot\left(1-e^{-\frac{\lambda P_{0} t}{2}}\right)\\
&+N \cdot \frac{\lambda\left(2-P_{0}\right)t}{2}\cdot e^{-\frac{\lambda P_{0}t}{2}}.
\end{aligned}
\end{equation}
To sum up, we obtain the differential equation for $Var\left[S_{A_{1}}(t)\right]$ as follows
\begin{equation}
\begin{aligned}
&\frac{d\left\{Var\left[S_{A_{1}}(t)\right]\right\}}{d t}\\
&=N \lambda \cdot\left(-3+\frac{3}{2} P_{0}\right) \cdot e^{-\frac{\lambda P_{0} t}{2}}+N \lambda \cdot\left(2-P_{0}\right) \cdot e^{-\lambda P_{0} t}\\
&+N \cdot \frac{\lambda^{2}\left(2-P_{0}\right) P_{0}}{2} \cdot t \cdot e^{-\frac{\lambda P_{0} t}{2}}+N \lambda
\end{aligned}
\end{equation}
Solving the above equation completes the proof
\begin{equation}
\begin{aligned}
&Var\left[S_{A_{1}}(t)\right]\\
&=\frac{N \lambda \cdot\left(-3+\frac{3}{2} P_{0}\right)}{\frac{1}{2} \lambda P_{0}}\cdot\left(1-e^{-\frac{\lambda P_{0} t}{2}}\right)+\frac{N \lambda \cdot\left(2-P_{0}\right)}{\lambda P_{0}} \cdot\left(1-e^{-\lambda P_{0} t}\right) \\
&+N \cdot \frac{2\left(2-P_{0}\right)}{P_{0}}\cdot\left(1-e^{-\frac{\lambda P_{0} t}{2}}\right)
-N \lambda t\cdot\left(2-P_{0}\right)\cdot e^{-\frac{\lambda P_{0} t}{2}}+N \lambda t \\
&=\frac{N\left(2-P_{0}\right)}{P_{0}}\cdot e^{-\frac{\lambda P_{0} t}{2}}\cdot\left(1-e^{-\frac{\lambda P_{0} t}{2}}\right)+N \lambda t \cdot\left[1-\left(2-P_{0}\right) \cdot e^{-\frac{\lambda P_{0} t}{2}}\right].
\end{aligned}
\end{equation}

\section{Proof of Theorem \ref{Thm4}}
\label{app4}

We first calculate $E\left[S_{A_{0}}(t)\right]$. Based on Eq. \eqref{CME3} we have
\begin{equation}
\begin{aligned}
&\frac{d\{E\left[S_{A_{0}}(t)\right]\}}{dt}\\
&=\sum_{x=0}^{N} x\cdot \frac{d\left\{P \left[S_{A_{0}}(t)=x\right]\right\}}{d t}=\sum_{x=1}^{N} x\cdot \frac{d\left\{P \left[S_{A_{0}}(t)=x\right]\right\}}{d t}\\
&=-\sum_{x=1}^{N} x\cdot \frac{\lambda\left(1-P_{0}\right)(N-x) x}{2 N}\cdot P\left[S_{A_{0}}(t)=x\right]\\
&-\sum_{x=1}^{N} x \cdot\frac{\lambda x\left[N-x\left(1-P_{0}\right)\right]}{2 N} \cdot P\left[S_{A_{0}}(t)=x\right]\\
&+\sum_{x=1}^{N} x\cdot \frac{\lambda\left(1-P_{0}\right)(N-x+1)(x-1)}{2 N}\cdot P\left[S_{A_{0}}(t)=x-1\right]\\
&+\sum_{x=1}^{N} x \cdot\frac{\lambda(x+1)\left[N-(x+1)\left(1-P_{0}\right)\right]}{2 N}\cdot P\left[S_{A_{0}}(t)=x+1\right]\\
&=-\frac{\lambda P_{0}}{2}\cdot E\left[S_{A_{0}}(t)\right].
\end{aligned}
\end{equation}
Its solution is given by
\begin{equation}
\begin{aligned}
E\left[S_{A_{0}}(t)\right]=N \cdot e^{-\frac{\lambda P_{0}}{2} t}.
\end{aligned}
\end{equation}
For $E\left[S^2_{A_{0}}(t)\right]$,
\begin{equation}
\begin{aligned}
&\frac{d\{E\left[S_{A_{0}}^{2}(t)\right]\}}{dt}\\
&=\sum_{x=0}^{N} x^{2}\cdot \frac{d\left\{P \left[ S_{A_{0}}(t)=x\right]\right\}}{d t}=\sum_{x=1}^{N} x^{2}\cdot \frac{d\left\{P \left[ S_{A_{0}}(t)=x\right]\right\}}{d t}\\
&=-\sum_{x=1}^{N-1} x^{2}\cdot \frac{\lambda\left(1-P_{0}\right)(N-x) x}{2 N}\cdot P\left[S_{A_{0}}(t)=x\right]\\
&-\sum_{x=1}^{N-1} x^{2}\cdot \frac{\lambda x\left[N-x\left(1-P_{0}\right)\right]}{2 N} \cdot P\left[S_{A_{0}}(t)=x\right]\\
&+\sum_{x=1}^{N-1} x^{2}\cdot \frac{\lambda\left(1-P_{0}\right)(N-x+1)(x-1)}{2 N}\cdot P\left[S_{A_{0}}(t)=x-1\right]\\
&+\sum_{x=1}^{N-1} x^{2}\cdot \frac{\lambda(x+1)\left[N-(x+1)\left(1-P_{0}\right)\right]}{2 N}\cdot P\left[S_{A_{0}}(t)=x+1\right]\\
&=-\left(\frac{N \lambda P_{0}+\lambda\left(1-P_{0}\right)}{N}\right)\cdot E\left[S_{A_{0}}^{2}(t)\right]+\frac{\lambda\left(2-P_{0}\right)}{2}\cdot E\left[S_{A_{0}}(t)\right].
\end{aligned}
\end{equation}
Its solution is given by
\begin{equation}
\begin{aligned}
&E\left[S_{A_{0}}^{2}(t)\right]\\
&=N^{2}\cdot e^{-\left[\frac{N \lambda P_{0}+\lambda\left(1-P_{0}\right)}{N}\right] \cdot t}\left\{\frac{\left(2-P_{0}\right)}{N P_{0}+2\left(1-P_{0}\right)}\cdot e^{\left[\frac{N \lambda P_{0}+2 \lambda\left(1-P_{0}\right)}{2 N}\right] \cdot t}+\frac{P_{0}(N-1)}{N P_{0}+2\left(1-P_{0}\right)}\right\}.
\end{aligned}
\end{equation}
Note that
\begin{equation}
\begin{aligned}
Var\left[S_{A_{0}}(t)\right]=E\left[S_{A_{0}}^{2}(t)\right]-\left(E\left[S_{A_{0}}(t)\right]\right)^{2},
\end{aligned}
\end{equation}
we have
\begin{equation}
\begin{aligned}
&Var\left[S_{A_{0}}(t)\right]\\
&=\frac{\left(2-P_{0}\right) N^{2}}{N P_{0}+2\left(1-P_{0}\right)}\cdot e^{\frac{-\lambda P_{0}t}{2}}+\frac{N^{2} P_{0}(N-1)}{N P_{0}+2\left(1-P_{0}\right)}\cdot e^{-\left[\frac{N \lambda_{0}+\lambda\left(1-P_{0}\right)}{N}\right] \cdot t}-N^{2}\cdot e^{-\lambda P_{0} t}
\end{aligned}
\end{equation}
Taking note of $S_{A_{1}}(t)=N-S_{A_{0}}(t)$, we obtain that
\begin{equation}
\begin{aligned}
E\left[S_{A_{1}}(t)\right]=N-N\cdot e^{-\frac{\lambda P_{0}t}{2}}
\end{aligned}
\end{equation}
and
\begin{equation}
\begin{aligned}
&Var\left[S_{A_{1}}(t)\right]\\
&=\frac{\left(2-P_{0}\right) N^{2}}{N P_{0}+2\left(1-P_{0}\right)}\cdot e^{\frac{-\lambda P_{0}t}{2}}+\frac{N^{2} P_{0}(N-1)}{N P_{0}+2\left(1-P_{0}\right)}\cdot e^{-\left[\frac{N \lambda P_{0}+\lambda\left(1-P_{0}\right)t}{N}\right]\cdot t}-N^{2}\cdot e^{-\lambda P_{0}t}.
\end{aligned}
\end{equation}

\section{Proof of inequality \eqref{inequ}}
\label{app5}

In order to prove $Var_m\left[S_{A_{0}}(t)\right]\geq Var_a\left[S_{A_{0}}(t)\right]$, it is equivalent to show
\begin{equation}
\begin{aligned}
Var_m\left[S_{A_{0}}(t)\right]&-Var_a\left[S_{A_{0}}(t)\right]=\left[\frac{\left(2-P_{0}\right) N^{2}}{N P_{0}+2\left(1-P_{0}\right)}-N\right] \cdot e^{\frac{-\lambda P_{0}t}{2} }\\
+&N(N-1)\cdot e^{-\lambda P_{0}  t}\cdot\left(\frac{N P_{0}}{N P_{0}+2\left(1-P_{0}\right)}\cdot e^{-\frac{\lambda\left(1-P_{0}\right) t}{N}}-1\right)\geq0
\end{aligned}
\end{equation}
We can rewrite above equation as
\begin{equation}
\begin{aligned}
Var_m\left[S_{A_{0}}(t)\right]-Var_a\left[S_{A_{0}}(t)\right]=\frac{\left(N^{2}-N\right)}{N P_{0}+2\left(1-P_{0}\right)} \cdot e^{\frac{-\lambda P_{0} t}{2}}\cdot g(t),
\end{aligned}
\end{equation}
where
\begin{equation}
\begin{aligned}
g(t)=\left(2-P_{0}\right)+N P_{0} \cdot e^{-\frac{\lambda\left(1-P_{0}\right ) t}{N}}\cdot e^{\frac{-\lambda P_{0} t}{2}}-\left(N P_{0}+2\left(1-P_{0}\right)\right)\cdot e^{\frac{-\lambda P_{0} t}{2}}.
\end{aligned}
\end{equation}
Note that $\frac{\left(N^{2}-N\right)}{N P_{0}+2\left(1-P_{0}\right)} \cdot e^{\frac{-\lambda P_{0} t}{2}}$ is always non-negative, it is sufficient to show $g(t)\geq 0$.
Consider the derivative of $g(t)$, we have
\begin{equation}
\begin{aligned}
&g^{\prime}(t)\\
&=\lambda N P_{0}\cdot\left(-\frac{\left(1-P_{0}\right)}{N}-\frac{P_{0}}{2}\right) \cdot e^{-\frac{\lambda\left(1-P_{0}\right) t}{N}}\cdot e^{\frac{-\lambda P_{0} t}{2}}+\frac{\lambda P_{0}}{2}\cdot\left(N P_{0}+2\left(1-P_{0}\right)\right) \cdot e^{\frac{-\lambda P_{0} t}{2}} \\
&=\lambda P_{0} \cdot e^{\frac{-\lambda P_{0}t }{2}}\cdot\left(-1+P_{0}-\frac{N P_{0}}{2}\right)\cdot \left( e^{-\frac{\lambda\left(1-P_{0}\right) t}{N}}-1\right)\geq 0
\end{aligned}
\end{equation}
Namely, $g(t)$ is monotonic increasing. Note that
\begin{equation}
\begin{aligned}
g(0)=\left(2-P_{0}\right)+N P_{0}-\left(N P_{0}+2-2 P_{0}\right)=P_{0} \geq 0,
\end{aligned}
\end{equation}
we have $g(t)\geq 0$, which completes the proof.

%


\end{document}